\documentclass[english,american,lengthcheck,final,prl,showpacs]{revtex4-1}
\usepackage[T1]{fontenc}
\usepackage[latin9]{inputenc}
\usepackage{color}
\usepackage{babel}
\usepackage{pdfpages}
\usepackage{amsmath}
\usepackage{graphicx}
\usepackage{esint}
\usepackage[unicode=true,pdfusetitle,
 bookmarks=true,bookmarksnumbered=false,bookmarksopen=false,
 breaklinks=false,pdfborder={0 0 1},backref=false,colorlinks=true]
 {hyperref}

\makeatletter

 \@ifundefined{textcolor}{}
 {%
   \definecolor{BLACK}{gray}{0}
   \definecolor{WHITE}{gray}{1}
   \definecolor{RED}{rgb}{1,0,0}
   \definecolor{GREEN}{rgb}{0,1,0}
   \definecolor{BLUE}{rgb}{0,0,1}
   \definecolor{CYAN}{cmyk}{1,0,0,0}
   \definecolor{MAGENTA}{cmyk}{0,1,0,0}
   \definecolor{YELLOW}{cmyk}{0,0,1,0}
 }

\makeatother

\begin{document}

\title{Coherent dynamical recoupling of diffusion-driven decoherence in
magnetic resonance}

\author{Gonzalo A. \'Alvarez}

\affiliation{Department of Chemical Physics, Weizmann Institute of Science, Rehovot,
76100, Israel}

\author{Noam Shemesh}

\affiliation{Department of Chemical Physics, Weizmann Institute of Science, Rehovot,
76100, Israel}

\author{Lucio Frydman}

\email{lucio.frydman@weizmann.ac.il}

\affiliation{Department of Chemical Physics, Weizmann Institute of Science, Rehovot,
76100, Israel}
\begin{abstract}
During recent years, dynamical decoupling (DD) has gained relevance
as a tool for manipulating and interrogating quantum systems. This
is particularly relevant for spins involved in nuclear magnetic resonance
(NMR), where DD sequences can be used to prolong quantum coherences,
or for selectively couple/decouple the effects imposed by random environmental
fluctuations. In this Letter, we show that these concepts can be exploited
to selectively recouple diffusion processes in restricted spaces.
The ensuing method provides a novel tool to measure restriction lengths
in confined systems such as capillaries, pores or cells. The principles
of this method for selectively recoupling diffusion-driven decoherence,
its standing within the context of diffusion NMR, extensions to the
characterization of other kinds of quantum fluctuations, and corroborating
experiments, are presented.
\end{abstract}

\pacs{03.65.Yz, 76.60.Es, 76.60.Lz, 82.56.Lz.}

\maketitle

\paragraph{Introduction.---}

Understanding and manipulating the lifetimes of quantum coherences,
are central goals of contemporary physics. Quantum decoherence can
be mitigated in several ways \cite{Zurek2003}; most often, this is
achieved by rotation pulses that decouple the system from its environment.
While such trains of refocusing pulses are known since the early days
of nuclear magnetic resonance (NMR) \cite{Hahn1950,Carr1954,*Meiboom1958,Abragam1961a},
these concepts have been generalized within the quantum information
community by ``dynamical decoupling'' (DD) ideas \cite{Viola1999,*Khodjasteh2005,Kofman2001b,Uhrig2007}.
These efforts aim at modulating the dephasing effects that environmental
fluctuations have on a quantum spin system; i.e., on filtering out
modes in the environment's spectral density noise. One form to achieve
this entails designing DD sequences so that the time modulations experienced
by the spins, will minimize their overlap with the the noise's spectral
density \cite{Kofman2001b,Uhrig2007,Cywinski2008,*Gordon2008,*Pasini2010,Ajoy2011}.
This is usually exploited to characterize an environment's spectral
density by varying the number of refocusing pulses or the interpulse
delays \cite{Meriles2010,*Almog2011,*Bylander2011,Alvarez2011}. Introducing
such changes, however, may introduce complications of their own: varying
the number of pulses may become a source of apparent decoherence via
pulse imperfections \cite{Alvarez2010c,*Ryan2010a,*Souza2011a,*Franzoni2012};
and even if pulses are kept constant, varying their interpulse delay
may lead to different total experimental times and hamper the measurement
being sought -for instance, by imparting differing spin-spin relaxation
($T_{2}$) weightings. These complications can be avoided if DD sequences
retain a constant overall duration and number of pulses \cite{Jenista2009,Smith2012},
but depart from the dogma of using constant inter-pulse delays \cite{Uhrig2007}.
In NMR this has been suggested as a new imaging (MRI) source of contrast
\cite{Jenista2009}. In spectroscopic characterizations, the power
of this concept was recently demonstrated by Selective-Dynamical-Recoupling
(SDR) sequences \cite{Smith2012}, where both the total evolution
time and the number of pulses remain fixed, while the interpulse delay
distribution is systematically varied. Unlike conventional CPMG sequences,
the SDR approach is immune to decoherence effects driven by cumulative
pulse imperfections and/or to intrinsic $T_{2}$ spin-spin relaxation.
SDR leads to a constant-time experiment with a fixed number of pulses,
that probes chemical identities via oscillatory modulations derived
from chemical shifts \cite{Smith2012}.

This Letter addresses a different kind of decoherence effects; namely,
those arising from spins diffusing in restricted spaces in the presence
of a magnetic field gradient. Seeking improved ways of characterizing
these phenomena has been a central theme in NMR and MRI \cite{Callaghan1991,*Callaghan2011,Price1997,*Sen2004,*Grebenkov2007},
and has enabled a wide range of studies ranging from oil prospecting,
to developmental brain studies -passing through many areas of physics,
chemistry and biology \cite{Basser1994b,*Song2008,*Laun2011,Kukla1996,*Kuchel1997,*Peled1999,*LeBihan2003,*Mair1999,*Song2000}.
As is shown here SDR can be a powerful approach to probe diffusion-related
phenomena, whilst filtering pulse imperfections and intrinsic $T_{2}$
decay effects. This form of DD can probe the spectral density of a
stochastic diffusion process, yielding information about the latter
and reflecting in a straightforward manner the restricting lengths
of the system. The ensuing approach turns out to be different from
typical modulated gradient sequences in that, rather than probing
a constant decay law that includes the diffusion spectrum \cite{Mitra1992,Callaghan1995,Stepisnik2006,Gore2010},
it probes how dynamics transition from free to restricted decay rates.
This allows one to probe even very small ($\sim\mu$m) length scales,
without requiring the very strong magnetic field gradients that are
conventionally needed to observe diffusion-diffraction phenomena \cite{Stejskal1965a,Callaghan1991,Shemesh2012}.
Furthermore, although this approach is illustrated here for NMR, it
is a conceptually general way to probe noise correlation times -in
particular those determining restricting length scales in complex
systems. 

To set the stage for this novel approach to monitor constrained diffusion,
we first provide a general analysis of random translation under dynamical
decoupling and its relationship with spectral density. We then analyze
the effects of free and restricted diffusion on the SDR modulations.
Finally, we experimentally demonstrate how these modulations can be
harnessed for accurately measuring compartment sizes under easily
amenable conditions, in a noninvasive manner.

\paragraph*{Modeling diffusion under dynamical decoupling.---}

Whereas the method proposed herein is general for probing fluctuations
in a quantum two-level system interacting with a bath \cite{Abragam1961a,Kofman2001b},
we consider for conciseness an ensemble of $S=1/2$ spins that do
not interact with each other, but are coupled to a classical external
magnetic field. This field involves a uniform component along the
$z$ axis defining a dominant Larmor frequency, and a perturbing linear
field gradient $G$. Due to this gradient, diffusion-induced displacements
will subject the spins to fluctuating precession frequencies. In a
usual rotating frame of reference \cite{Abragam1961a}, the resulting
Hamiltonian will be a pure dephasing interaction $\widehat{\mathcal{H}}_{SE}\left(t\right)=\omega_{SE}\left(t\right)\hat{S}_{z}$,
where $\omega_{SE}\left(t\right)=\gamma G\, r(t)$ is a the frequency
(noise) felt by the spin, with $r$ denoting the position of the diffusing
spin along the field gradient direction $G=\partial B_{z}/\partial r$. 

\begin{figure}
\includegraphics[bb=0bp 0bp 354bp 202bp,width=1\columnwidth]{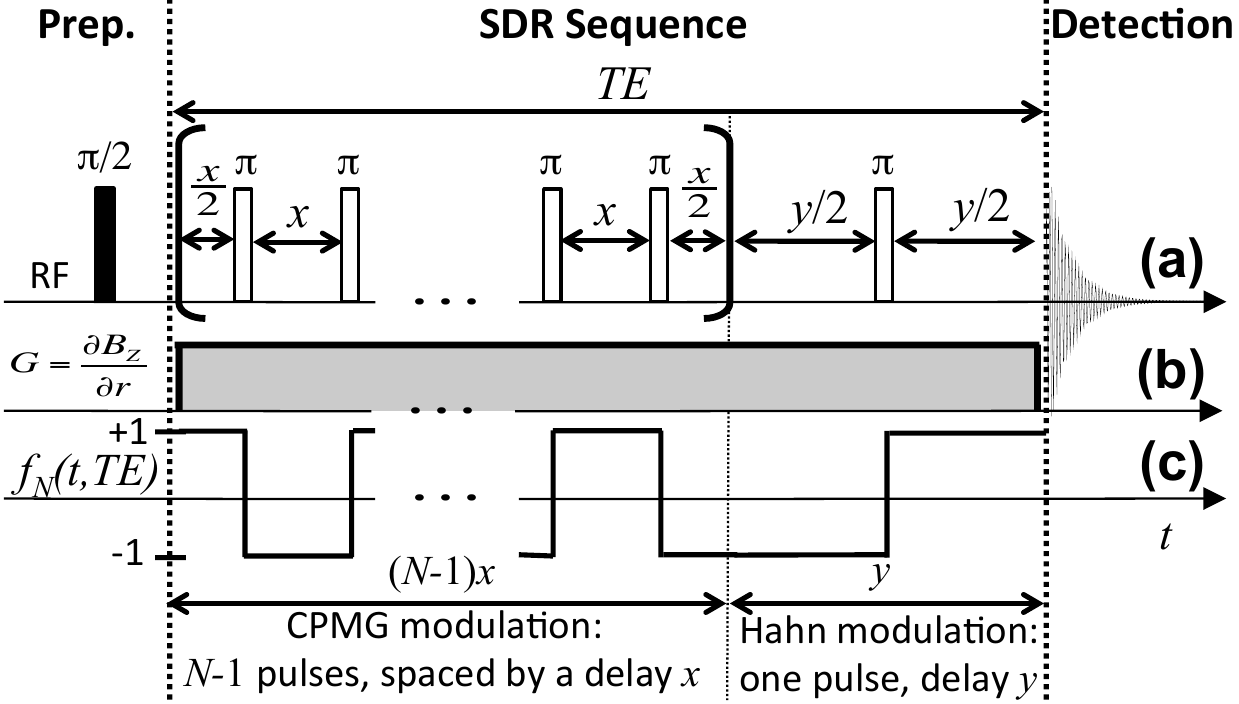}

\caption{Selective dynamical recoupling (SDR) sequence proposed for probing
the diffusion spectrum, and involving (a) a sequence of $N$ rf $\pi$
pulses applied to the spins during a total evolution time $TE$; and
(b) a constant magnetic field gradient $G$. A conventional CPMG sequence
would arise if $x=y=TE/N$; we refer to the $N=1$, $x=0$ case as
a Hahn-echo sequence. (c) Modulating function $f_{N}\left(t\right)$
imposed by the sequence of pulses. }
\label{Flo:scheme}
\end{figure}

Consider the application of a sequence of strong $\pi$ pulses as
shown in Fig. \ref{Flo:scheme}a, that periodically refocuses the
spin ensemble after it has been subject to excitation. 
The sequence assumes $N$ instantaneous pulses at times $t_{i}$,
with an initial delay $t_{1}-t_{0}=x/2$ ($t_{0}=0$), uniform delays
$t_{i}-t_{i-1}=x$ between the pulses for $i=2,..,N-1$, and a final
pulse at $t_{N}=TE-y/2$. $t_{N+1}=TE$ is the total evolution time,
and the $x$ and $y$ delays are such that $TE=y+(N-1)x$. Given the
equidistant train of $\pi$ pulses involved in the first part of the
sequence we refer to it as involving a ``CPMG'' modulation \cite{Carr1954,Meiboom1958},
and to the final single-inversion part of the sequence as a Hahn modulation
\cite{Hahn1950}. This conforms to the SDR sequence \cite{Smith2012}
shown in Fig. \ref{Flo:scheme}. A constant gradient $G$ given by
an external action or local fields, is assumed to be active throughout
the pulse train.

Under pulse-free conditions, the spin evolution operator for a given
realization of a spatial random walk will be $\exp\left\{ -i\phi(TE)\hat{S}_{z}\right\} $,
where $\phi(TE)$ is the accumulated phase gained by the diffusing
spin during $TE$. The effects that the pulse train in Fig. \ref{Flo:scheme}
will impose on the evolving spin can be accounted for by instantaneous
sign changes of the evolution frequencies $\omega_{SE}(t)$. After
applying the $N$ pulses the accumulated phase will be $\phi(TE)=\int_{0}^{TE}dt^{\prime}f_{N}(t^{\prime},TE)\omega_{SE}(t^{\prime})$,
where the modulating function $f_{N}(t^{\prime},TE)$ switches between
$\pm1$ as shown in Fig. \ref{Flo:scheme}c. Given an initial state
$\hat{\rho}_{0}=\hat{S}_{x}$, the normalized magnetization arising
from an ensemble of non-interacting and equivalent spins under the
effects of this sequence will be $M(TE)=\left\langle e^{-i\phi\left(TE\right)}\right\rangle $,
where the brackets account for an ensemble average over the random
phases. Without pulses $\left\langle \phi\left(TE\right)\right\rangle $
would depend on the position of each spin in the sample; as with all
DD sequences, however, the average for the SDR case will be $\left\langle \phi\left(TE\right)\right\rangle =0$.
Thus, assuming that the random phase $\phi\left(t\right)$ has a Gaussian
distribution \cite{Klauder1962,Stepisnik1999}, $M(TE)=\exp\left\{ -\frac{1}{2}\left\langle \phi^{2}(TE)\right\rangle \right\} $:
the signal will evidence a decay depending on the random phase's variance.
This argument is solely given by the spins' diffusion within $G$,
and can be written in a Fourier transform representation \cite{Callaghan1995,Kofman2001b,Stepisnik2006}
as:
\begin{equation}
\frac{1}{2}\left\langle \phi^{2}(TE)\right\rangle =\tfrac{\Delta\omega_{SE}^{2}}{2}\int_{-\infty}^{\infty}d\omega S(\omega)\left|F(\omega,TE)\right|^{2}.\label{eq:Mtenuestra}
\end{equation}
 This expression entails a product of the spectral density $\Delta\omega_{SE}^{2}\, S(\omega)$
characterizing the diffusion-driven fluctuation, times the filter
function $F(\omega,TE)$ given by the Fourier transform of the modulation
function $\sqrt{2\pi}f_{N}(t^{\prime},TE)$. The spectral density
$\Delta\omega_{SE}^{2}\, S(\omega)$ is given in turn by the Fourier
transform of the auto-correlation function $g\left(\tau\right)=\left\langle \Delta\omega_{SE}(t)\Delta\omega_{SE}(t+\tau)\right\rangle $,
where $\Delta\omega_{SE}(t)=\gamma G\,\left[r\left(t\right)-\left\langle r\left(t\right)\right\rangle \right]$
is the spin's instantaneous frequency deviation from its average value,
and $\Delta\omega_{SE}^{2}=\left\langle \Delta\omega_{SE}^{2}(0)\right\rangle $.
Assuming $g\left(\tau\right)$ follows an exponential decay, the spectral
density of this fluctuation will be given by the Lorentzian function
\cite{Klauder1962,stepisnik-multiplelorentzian} 
\begin{eqnarray}
\frac{\mathcal{FT}\left\{ g(\tau)\right\} }{\sqrt{2\pi}}=\Delta\omega_{SE}^{2}\, S(\omega) & = & {\frac{\Delta\omega_{SE}^{2}{\it \tau_{c}}}{\left(1+{\omega}^{2}\tau_{c}^{2}\right)\pi}}.
\end{eqnarray}
Here the correlation time $\tau_{c}$ will be associated to a characteristic
length $l_{c}$, given by the diffusion process according to the Einstein's
expression $l_{c}^{2}=2D_{0}\tau_{c}$, 
\begin{figure}
\includegraphics[width=1\columnwidth]{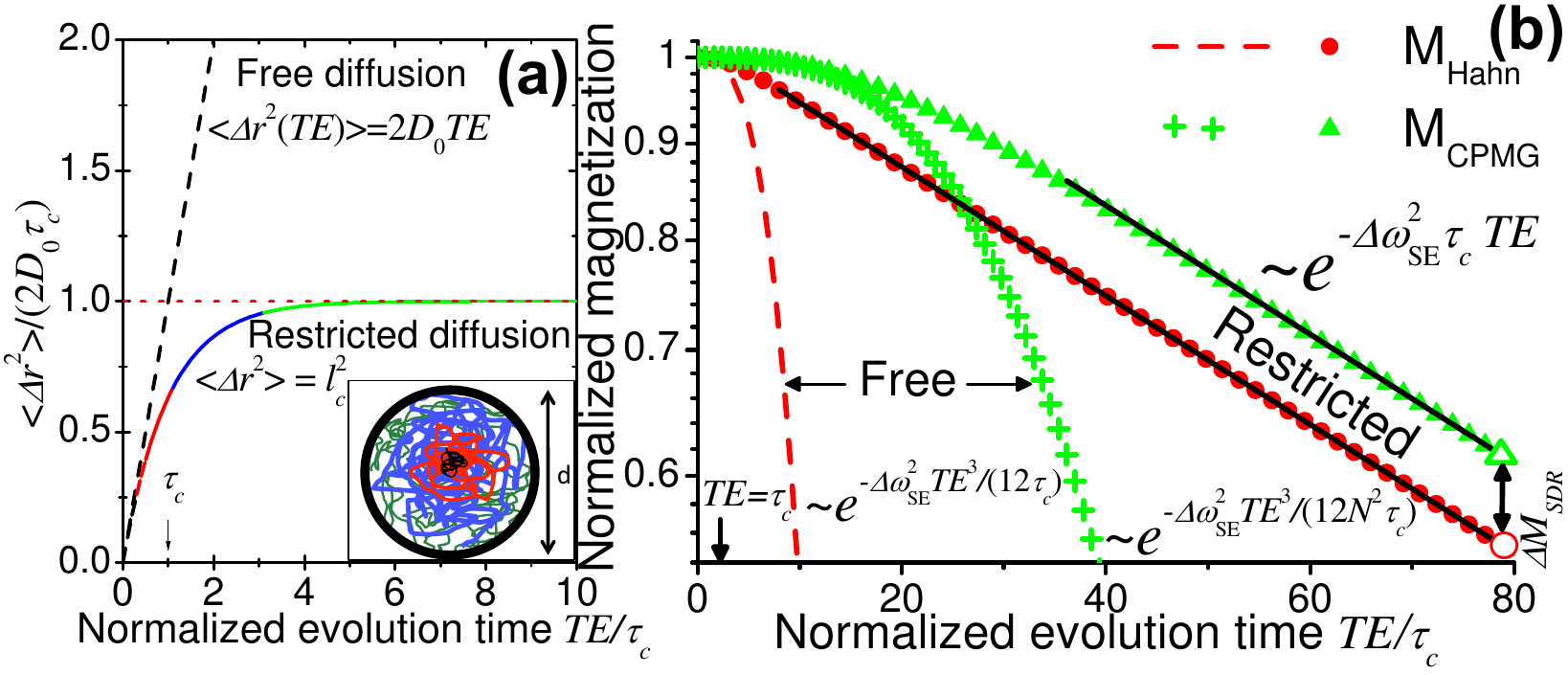}\caption{\label{fig:displacement_time_dependence}(Color online) (a) Normalized
mean square displacement (solid multi-color line) of the diffusing
spins in a restricted space. The dashed black line gives the free
diffusion law while the dotted gives the restriction length $l_{c}$.
The inset shows in different colors random trajectories for different
total times within a cylinder of diameter $d$, where $l_{c}\approx0.37d$.
The same color tones are in the solid line to show the different time
regimes of the spin trajectories. \label{fig:DD_decays}(b) Time evolution
of the spin magnetization under CPMG ($N=8$ pulses) and Hahn-echo
sequences for spins diffusing in a restricted space (triangles, circles),
and under free diffusion (crosses, dashes). The solid black lines
show the time range where the restricted diffusion effects dominate;
the difference $\Delta M_{SDR}$ between these lines gives a contrast,
over which signals can be coherently modulated by a suitable SDR filter
function. }
\end{figure}
where $D_{0}$ is the free diffusion coefficient. It also follows
that $\Delta\omega_{SE}^{2}=\gamma^{2}G^{2}D_{0}\tau_{c}$. If considering
now diffusion in a pore or restricted cavity, the specific relation
between $l_{c}$ and the restriction length $d$ of the pore will
depend on its geometry; e.g. for cylinders a good approximation is
$\tau_{c}\approx0.26^{2}d^{2}/D_{0}$ \cite{Stepisnik1993,Stepisnik2006}
and then $l_{c}\approx0.37d$, where $d$ is the cylinder's diameter.
Figure \ref{fig:displacement_time_dependence} compares the different
behavior that, in units of the correlation time $\tau_{c}$, will
be evidenced by the spin's mean displacement $\left\langle \Delta r(t)^{2}\right\rangle =\left\langle \left[r(t)-r(0)\right]^{2}\right\rangle $
depending on whether diffusion is free or restricted.

\paragraph{Restricted and free diffusion: Effects on the SDR modulations.---}

With this scenario as background, we consider the effects of the sequence
in Fig. \ref{Flo:scheme} for probing the behaviors illustrated in
Fig. \ref{fig:displacement_time_dependence}a. The ``gist'' of SDR
is that it manages to distinguish these cases without varying $TE$
or the total number of intervening pulses, but rather using the flexibility
that the delays $x$ and $y$ in Fig. \ref{Flo:scheme}, afford for
shaping the $F(\omega,TE)$ filter function. To see this more clearly,
consider the two segments in the SDR sequence -the Hahn and the CPMG
modulations- separately. The diffusion-driven signal decay for a Hahn-echo
sequence \cite{Hahn1950} can be obtained analytically \cite{Kennan1994,SI-diffusionSDR};
its decay is shown in Fig. \ref{fig:DD_decays}b by the red dashes
and circles. Also the analytical expression for the signal decay of
a CPMG sequence, characterized by $N$ equispaced pulses ($x=y=TE/N$
in Fig. \ref{Flo:scheme}) can be calculated \cite{SI-diffusionSDR};
the ensuing magnetization decay is plotted in Fig. \ref{fig:DD_decays}b
(green crosses and triangles). The free diffusion regime exhibits
the simplest behavior: since the delays between pulses $x,y\ll\tau_{c}$,
the filter function $F$ peaks at frequencies $\omega\gg1/\tau_{c}$
\cite{Ajoy2011,Alvarez2011} and decoherence effects are dominated
by the tail of the spectral density $S(\omega)\propto{\frac{1}{{\omega}^{2}\tau_{c}\pi}}$
\cite{Pfitsch1999}. The signal decay therefore follows a decay rate
proportional to $ $$\omega^{-2}=(TE/N)^{2}$ (dashes and crosses
in Fig. \ref{fig:DD_decays}b). This result is derived in the original
CPMG paper for freely diffusing spins \cite{Carr1954}. By contrast,
in the restricted diffusion regime, $\tau_{c}$ is short due to the
confinement. The delays between pulses $x,y\gg\tau_{c}$, and the
dominant peaks of the filter functions $F$ are at frequencies $\omega\ll1/\tau_{c}$
\cite{Ajoy2011,Alvarez2011}. In these cases the signal decay follows
an exponential argument 
\begin{eqnarray}
\frac{1}{2}\left\langle \phi^{2}(TE)\right\rangle  & \approx & \Delta\omega_{SE}^{2}\tau_{c}\left(TE-\left(1+2N\right)\tau_{c}\right).\label{SCFGRdecays3}
\end{eqnarray}
 The exponential magnetization decay at a rate $\Delta\omega_{SE}^{2}\tau_{c}$,
is evidenced by the slopes in the solid black lines in Fig. \ref{fig:DD_decays}b.
The second term in Eq. (\ref{SCFGRdecays3}) \cite{Fiori2006footnote}
gives a shift depending on $N$, and is responsible for the $\Delta M_{SDR}$
gap separating the Hahn and the CPMG decays in Fig. \ref{fig:DD_decays}b.
While normally the usual expression used for the restricted diffusion
decay rate is just the first term of (\ref{SCFGRdecays3}) \cite{Kennan1994},
the second term derived here is unique to the SDR sequence and provides
a new degree of freedom for probing restrictions according to the
choice of $x$. In particular if $x\ll\tau_{c}\ll y$, the decay of
the signal during the SDR is dominated by the Hahn portion of the
sequence and approaches $M_{Hahn}^{restricted}(y)=\exp\left\{ 3\Delta\omega_{SE}^{2}\tau_{c}^{2}\right\} \exp\left\{ -\Delta\omega_{SE}^{2}\tau_{c}y\right\} $;
but if $x=y=TE/N\gg\tau_{c}$ the SDR decay will be $M_{CPMG}^{restricted}(TE,N)=$
$\exp\left\{ \left(1+2N\right)\Delta\omega_{SE}^{2}\tau_{c}^{2}\right\} $
$\exp\left\{ -\Delta\omega_{SE}^{2}\tau_{c}TE\right\} $. Thus, the
SDR approach allows one to probe $\tau_{c}$ -and hence a confinement
length $l_{c}$- from the difference between the Hahn and the CPMG
decays ($\Delta M_{SDR}$) that are built into the sequence. Notice
that if $TE,y\gg\tau_{c}$, $\Delta M_{SDR}/M_{Hahn}^{restricted}(TE)=\exp\left\{ 2\left(N-1\right)\Delta\omega_{SE}^{2}\tau_{c}^{2}\right\} -1$
independently from $TE,\, x$ or $y$. Moreover, while the exponential
rate typically used for determining $l_{c}$ is $\propto\Delta\omega_{SE}^{2}\tau_{c}\propto l_{c}^{4}$,
the shift term $\ln\left\{ \Delta M_{SDR}/M_{Hahn}^{restricted}(TE)+1\right\} =2\left(N-1\right)\Delta\omega_{SE}^{2}\tau_{c}^{2}=\left(N-1\right)l_{c}^{6}\gamma^{2}G^{2}/(4D_{0}^{2})$
amplifies this new source of contrast with $N$, and makes it a more
sensitive reporter on the value of $l_{c}$ as it is $\propto l_{c}^{6}$.

It follows that the sequence in Fig. \ref{Flo:scheme} can interrogate
restricted diffusion while fixing $TE$ as well as the number of inversion
pulses, by dividing an echo train into periods involving different
interpulse delays $x$ and $y$. By controlling the ratio $x/y$ one
can probe the spectral density $S(\omega)$ and determine the transition
between Hahn- and CPMG-dominated regimes. The total sequence's time
modulation $f^{SDR}$ will then be given by 
\begin{multline}
f_{N,x,y}^{SDR}\left(t,TE\right)=f_{N-1}^{CPMG}\left(t,(N-1)x\right)\\
+(-1)^{N-1}\, f_{1}^{Hahn}\left(t-\left(N-1\right)x,y\right),
\end{multline}
where $f_{N-1}^{CPMG}$ and $f_{1}^{Hahn}$ are the CPMG and the Hahn
modulating functions. The filter function $\left|F_{N,x,y}^{SDR}(\omega,TE)\right|^{2}$
associated to SDR will thus be the sum of a CPMG portion, a Hahn portion,
plus a cross term representing an interference between these two filters:
\begin{multline}
\left|F_{N,x,y}^{SDR}(\omega,TE)\right|^{2}=\left|F_{N-1}^{CPMG}\left(\omega,\left(N-1\right)x\right)\right|^{2}\\
+\left|F_{1}^{Hahn}\left(\omega,y\right)\right|^{2}+(-1)^{N-1}2\,\mathrm{Re}\left\{ e^{i\omega\left(TE-y\right)}\right.\\
\left.F_{N-1}^{CPMG}\left(\omega,\left(N-1\right)x\right)\overline{F_{1}^{Hahn}\left(\omega,y\right)}\right\} .\label{eq:SDR_filter}
\end{multline}
This filter-function formalism allows one to derive a solution for
the resulting signal decay 
\begin{multline}
M_{SDR}(TE,x,y,N)=M_{CPMG}\left(\left(N-1\right)x,N-1\right)\\
\times M_{\mathrm{Hahn}}\left(y\right)\times M_{Cross-SDR}(TE,x,y,N),\label{eq:SDRsignal_decay}
\end{multline}
whose analytical expression is given in the supplementary information
\cite{SI-diffusionSDR}.

It is worth concluding this paragraph by noting that the correlation
time $\tau_{c}$ can also be extracted by comparing the exponential
decay curves of independent Hahn and CPMG sequences, or by changing
the $N/TE$ ratio of a CPMG set. Such variations, however, would require
comparing signal decays arising from measurements involving different
number of pulses or different overall $TE$'s. Only SDR manages to
keep those parameters -whose variation could eclipse the diffusion
effects being sought constant throughout the measurements.

\paragraph{SDR measurements of restriction lengths.---}

\begin{figure}
\includegraphics[width=0.8\columnwidth]{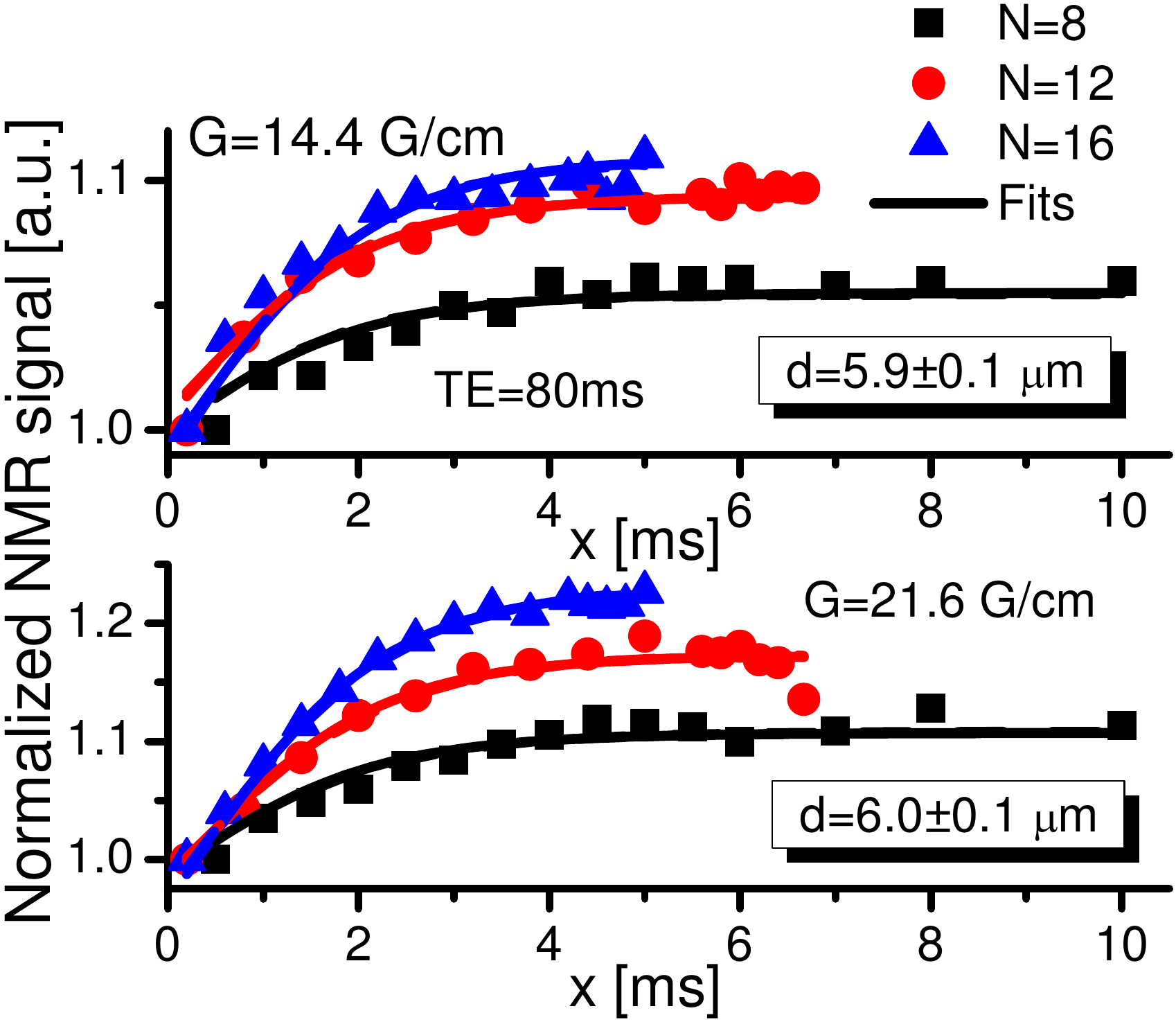}

\caption{\label{fig:SDR_exp}(Color online) Experimental SDR signals normalized
with the first data point (symbols) as a function of the $x$ delays.
The solid lines are analytical fittings of Eq. (\ref{eq:SDRsignal_decay})
to the experimental curve. By using the measured diffusion coefficient
$D_{0}\sim2.3\times10^{-5}cm^{2}/s$, the fitted diameter $d$ given
in the plots is in agreement with the nominal value $d=5\pm1\mu$m.}
\end{figure}

As proof of SDR's capabilities to accurately measure restricted diffusion,
the sequence was applied to examine the diameter of water-filled microcapillaries
with a nominal value of $5\pm1\,\mu$m (Polymicro Technologies, Phoenix,
Az, USA). A free diffusion coefficient $D_{0}\sim2.3\times10^{-5}$cm$^{2}$/s
was measured by a conventional NMR sequence \cite{Stejskal1965a}
in which the orientation of an applied gradient coincided with the
principal axis of the microcapillaries. $^{1}$H SDR curves of water
diffusing within the capillaries were recorded in the presence of
a transverse magnetic field gradient using a 9.4T Bruker microimaging
NMR scanner, where the effects of background gradients ($G=0$) were
found negligible. Figure \ref{fig:SDR_exp} shows the SDR modulations
observed as a function of $x$ with $TE=80$ms, for values of $G=$14.4
and 21.6 G/cm. A transition from the diffusion-driven Hahn decay ($x\sim0$)
to the CPMG decay ($x=TE/N$) can be clearly appreciated in each data
set; the difference between the $x=0$ and $x=TE/N$ conditions, $\Delta M_{SDR}$,
together with the dependence on $x$ in general, provide a robust
determination of the diffusion's correlation time -and from there
of $l_{c}$. The fit between the analytical expression derived for
the SDR decay (Eqs. (S.5)-(S.22) in the Sup. Inf. \cite{SI-diffusionSDR})
corresponding to particles diffusing in a cylinder \cite{Stepisnik1993,Stepisnik2006}
and the experimental data is excellent, and so is the agreement with
the nominal inner diameter provided by the capillaries' supplier.
Note the resemblance in the behavior of the SDR curves in Fig. \ref{fig:SDR_exp}
and the $\left\langle \Delta r^{2}\right\rangle $ in Fig. \ref{fig:displacement_time_dependence}a:
in both cases curves plateau for times $x>\tau_{c}$, evidence a full
sampling of the restricting space.

\paragraph{Discussion.---}

The fact that the different $\Delta M_{SDR}$ measured by SDR at constant
$TE$ and $N$ are solely defined by $\tau_{c}$, provides a novel
and simpler approach for determining restriction lengths $l_{c}$
by NMR. Alternative noninvasive methodologies for probing the compartment
dimensions, foremost among them diffusion-diffraction phenomena \cite{Callaghan1991,Stejskal1965a,Shemesh2012},
require very strong magnetic field gradients - stronger by $\sim2$
orders of magnitude than the gradients demanded by SDR, when small
pores are considered. For example, measuring diffraction patterns
in cylindrical pores characterized by a restricting length scale of
$\sim5\mu$m such as the ones used in this study, would require gradient
amplitudes exceeding 1000 G/cm. Furthermore, methodologies that focus
on probing a transition from free to restricted diffusion, will usually
do so focusing on the deviations observed for the spectral density
from power-law tails \cite{Mitra1992,Stepisnik2006,Gore2010}. Instead,
in the SDR case, the decay is dominated by the restricted diffusion
regime: this makes $\Delta M_{SDR}$ a much more robust and sensitive
means for determining length constraints. The resolvable restricting
sizes of the ensuing method will eventually depend on the ratio $\Delta M_{SDR}/M_{Hahn}^{restricted}(TE)$
being larger than the signal to noise ratio; among the factors that
can magnify this ratio count $N$ and $G$, which enhance $\Delta M_{SDR}$
as $\left(N-1\right)l_{c}^{6}\gamma^{2}G^{2}/(4D_{0}^{2})$. Probing
restricted diffusion in tissues for example, where the diffusion coefficent
$D_{0}\sim0.7\times10^{-5}$ cm$^{2}$/s, will lead to variations
in $\Delta M_{SDR}/M_{Hahn}^{restricted}(TE)$ of between 10\% - 250\%
for compartment in the 1-1.5$\mu$m range and typical gradient amplitudes
of $\sim50$ G/cm and $N=16$. If stronger gradients (>500 G/cm) and
spin-abundant porous systems are considered, sizes on the order of
hundreds of nanometers should become clearly resolvable (depending
on the intrinsic $D_{0}$, which can also be controlled via temperature).
As $N$ increases the sensitivity of the measurements also grows and
the detectable size limits may be reduced even further. However, SDR's
robustness will also depend on the accuracy of the refocusing pulses.
Furthermore, SDR may be biased towards longer $T_{2}$ species, as
a result of its constant-time nature. 

This study demonstrated another instance where -as was the case with
chemical exchange and $J$-coupling effects \cite{Smith2012}- suitable
DD schemes can extract coherent modulations from restricted NMR spectral
fluctuations. As in previous spectroscopic demonstrations, a key ingredient
to achieve these modulations is to have spins exchanging within a
discrete/bound frequency spectrum, which DD can then probe by adjusting
its filtering characteristics. The ensuing SDR method is particularly
simple for determining the restricting length scales in complex, opaque
systems \cite{Callaghan1991,Kukla1996,*Kuchel1997,*Peled1999,*LeBihan2003,*Mair1999,*Song2000}
where the pore topology governs physical or biological properties
of the materials. Applications of SDR to probe other kind of constrained
or pinned diffusive processes like charges diffusing in conducting
crystals \cite{Feintuch2004} or spin diffusion in molecules \cite{Bloembergen1949,*Suter1985}
can also be envisaged. Additionally, we expect that this method can
be useful for imaging other kinds of spectra at the nanoscale; for
example by sensing the correlation times of noise fluctuation generated
by a host system on single spins in diamonds \cite{Zhao2011,*Staudacher2013,*Steinert2013}. 
\begin{acknowledgments}
We are grateful to Pieter Smith, Guy Bensky and Gershon Kurizki (Weizmann
Institute) for fruitful discussions and to Prof. Yoram Cohen (Tel
Aviv University) for providing the micro-capillaries used in this
study. This research was supported by a Helen and Martin Kimmel Award
for Innovative Investigation, and the generosity of the Perlman Family
Foundation. GAA acknowledges the support of the European Commission
under the Marie Curie Intra-European Fellowship for career Development
Grant No. PIEF-GA-2012-328605.
\end{acknowledgments}
\bibliographystyle{apsrev4-1}
\bibliography{/Users/galvarez/Dropbox/Papers/BibTeX/bibliography}

\selectlanguage{english}%
\pagebreak{}

\begin{widetext}

\pagebreak{}.



\section*{Supplementary material}

# Supplementary information for "Coherent dynamical recoupling of diffusion-driven decoherence in magnetic resonance"
## Full analytical expressions of the SDR signal evolution under diffusion

Gonzalo A. Álvarez, Noam Shemesh, and Lucio Frydman

*Department of Chemical Physics, Weizmann Institute of Science, Rehovot, 76100, Israel*

As mentioned in the section "Modeling diffusion under dynamical decoupling" section of the main text, the normalized magnetization arising from an ensemble of non-interacting and equivalent spins under the effects of a sequence of pulses will be $M(t) = \langle e^{-i\phi(t)} \rangle$, where the brackets account for an ensemble average over the random phases $\phi(t)$. For the dynamical decoupling sequences being considered the average phase $\langle \phi(t) \rangle$ will be equal to zero. Then, assuming that the random phase $\phi(t)$ has a Gaussian distribution [1, 2], $M(t) = \exp\left\{-\frac{1}{2}\langle \phi^2(t) \rangle\right\}$: the signal will evidence a decay depending on the random phase's variance. It is convenient to describe this variance in terms of the modulating function $f_N(t', TE)$ introduced in Fig. 1:

$$\frac{1}{2}\langle \phi^2(TE) \rangle = \tfrac{1}{2}\int_0^{TE} dt' \int_0^{TE} dt'' f_N(t', TE) f_N(t'', TE) \langle \omega_{SE}(t')\omega_{SE}(t'') \rangle \tag{S.1}$$

$$= \tfrac{1}{2}\int_{-\infty}^{\infty} dt' \int_{-\infty}^{\infty} dt'' f_N(t', TE) f_N(t'', TE) g(t'' - t'), \tag{S.2}$$

where $f_N(t', TE) = 0$ if $t' < 0$ or $t' > TE$ (i.e., outside the evolution time range), and the evolution is given in terms of the frequency fluctuation correlation function $g(\tau) = \langle \Delta\omega_{SE}(t')\Delta\omega_{SE}(t'+\tau) \rangle$. This correlation function is related to the fluctuation's spectral density $S(\omega)$ by a Fourier transform: $\mathcal{FT}\{g(\tau)\}/\sqrt{2\pi} = \Delta\omega_{SE}^2 S(\omega)$, where $\Delta\omega_{SE}^2 = \langle \Delta\omega_{SE}^2(0) \rangle$ is the mean square frequency fluctuation. Equation (S.1) can thus be recast in its Fourier representation [3–8] as:

$$\frac{1}{2}\langle \phi^2(TE) \rangle = \tfrac{\Delta\omega_{SE}^2}{2} \int_{-\infty}^{\infty} d\omega S(\omega) |F(\omega, TE)|^2, \tag{S.3}$$

where $F(\omega, TE)$ is the filter function introduced in Eq. (1) of the main text. Under the usual assumption of an exponential correlation function $g(\tau) = \Delta\omega_{SE}^2 \exp(-\tau/\tau_c)$, the ensuing spectral density will be

$$\Delta\omega_{SE}^2 S(\omega) = \frac{\Delta\omega_{SE}^2 \tau_c}{(1 + \omega^2 \tau_c^2)\pi}, \tag{S.4}$$

where $\tau_c$ is the correlation time of the fluctuations[1].

To calculate now the filter function $F(\omega, TE)$ for the SDR sequence, it is convenient to consider the three time-modulating functions $f(t, TE)$ in Fig. S.1. $f^{free}$ is a Boxcar function, null if $t < 0$ or $t > TE$ and constant otherwise (Fig. S.1a). Its filter function is therefore

$$F^{free}(\omega, TE) = \sqrt{2\pi}\mathcal{FT}\{f_{free}(t, TE)\} = e^{-\frac{i\omega TE}{2}} \frac{\sin(\omega TE/2)}{\omega/2}. \tag{S.5}$$

Following the recipe in Eqs. (S.1) and (S.3), this filter leads to a magnetization [10, 11]

$$M_{free}(TE) = \exp\left\{-\Delta\omega_{SE}^2 \tau_c TE \left[1 - \frac{\tau_c}{TE}\left(1 - \exp\left(-\frac{TE}{\tau_c}\right)\right)\right]\right\}. \tag{S.6}$$

Similarly one can obtain the Hahn echo filter by FT of the step function in Fig. S.1b:

$$F_1^{Hahn}(\omega, TE) = ie^{-\frac{i\omega TE}{2}} \frac{4\sin^2\left(\frac{\omega TE}{4}\right)}{\omega}, \tag{S.7}$$

---

[1] For a constant $G$ and well defined pore geometries, the spectral density is a weighted sum of Lorentzian functions with different correlation times [7, 9]. However, usually one of these functions is more significant than the others and, particularly in our method dealing with confined planar, spherical or cylindrical geometries, a single term of the kind given in (S.4) dominates the SDR evolution. If the remaining terms also need to be accounted for, a weighted sum of decay terms in $\frac{1}{2}\langle \phi^2(TE) \rangle$ with suitable, multiple values for $\Delta\omega_{SE,i}^2$ and $\tau_{c,i}$ in a series of Lorentzian terms need to be considered.

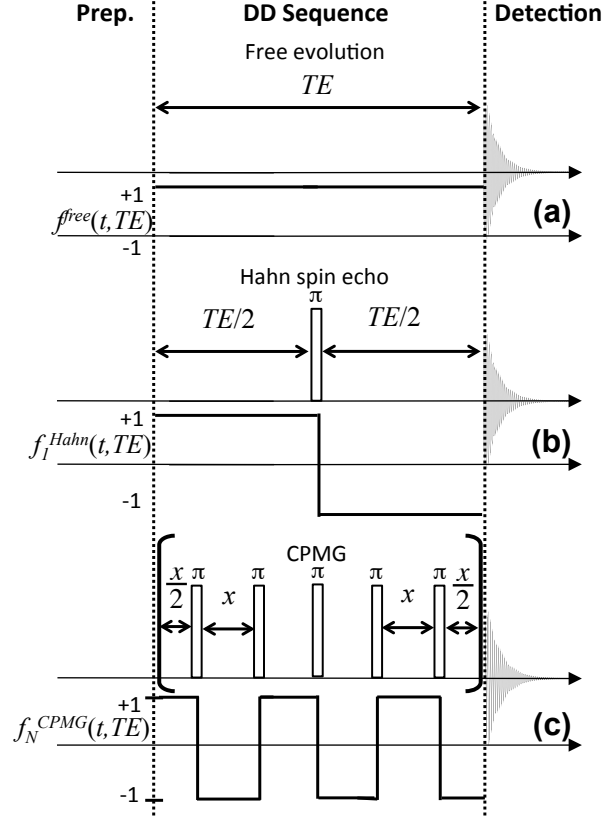

Figure S.1. NMR sequences and its modulation function. The empty squares are $\pi$ pulses. (a) Free evolution sequence, i.e. no pulses are applied and the modulation function $f^{free}(t,TE)$ is just a Boxcar function. (b) Hahn spin echo sequence and its modulationg function $f_1^{Hahn}(t,TE)$. (c) CPMG sequence with $N=5$ pulses, where $x=TE/N$, and its modulating function $f_N^{CPMG}(t,TE)$.

and the corresponding signal becomes

$$M_{\text{Hahn}}(TE) = \exp\left\{-\Delta\omega_{SE}^2 \tau_c TE\left[1 - \frac{\tau_c}{TE}\left(3 + \exp\left(-\frac{TE}{\tau_c}\right) - 4\exp\left(-\frac{TE}{2\tau_c}\right)\right)\right]\right\}, \tag{S.8}$$

as is known from Refs. [10, 11]. Finally, for the CPMG square-wave modulation function in Fig. S.1c, the corresponding filter can be written as

$$F_N^{CPMG}(\omega, TE) = \frac{i2e^{-\frac{i\omega TE}{2}}\sin^2\left(\frac{1}{4}\omega\frac{TE}{N}\right)\left(e^{\frac{1}{2}i\omega TE} + (-1)^{N+1}e^{-\frac{1}{2}i\omega TE}\right)}{\cos\left(\frac{1}{2}\omega\frac{TE}{N}\right)\omega}, \tag{S.9}$$

and we have calculated the corresponding magnetization as

$$M_{CPMG}(TE, N) = \exp\left\{-\Delta\omega_{SE}^2 \tau_c\left[TE - \tau_c(A+B)\right]\right\},$$
$$A = (2N+1) - (-1)^N e^{-\frac{TE}{\tau_c}},$$
$$B = -4\frac{(-1)^{N+1}e^{-\frac{TE}{\tau_c}}\left(e^{-\frac{3}{2}\frac{TE}{N\tau_c}} + e^{-\frac{1}{2}\frac{TE}{N\tau_c}} - e^{-\frac{TE}{N\tau_c}}\right) + e^{-\frac{3}{2}\frac{TE}{N\tau_c}} + e^{-\frac{1}{2}\frac{TE}{N\tau_c}} + e^{-2\frac{TE}{N\tau_c}}N + e^{-\frac{TE}{N\tau_c}}(N-1)}{\left(e^{-\frac{TE}{N\tau_c}} + 1\right)^2}. \tag{S.10}$$

One can use a similar approach to derive analytical expressions for the magnetization's decay incurred by the SDR sequence considered in our study. The modulating function for the sequence $(x/2 - \pi - x - \pi - x/2)^{N-1} - (y/2 - \pi - y/2)$, entails a sum of the two cases just analized

$$f_{N,x,y}^{SDR}(t, TE) = f_{N-1}^{CPMG}(t, (N-1)x) + (-1)^{N-1}f_1^{Hahn}(t - (N-1)x, y). \tag{S.11}$$

Therefore, its corresponding filter function is

$$\left|F_{N,x,y}^{SDR}(\omega,TE)\right|^2 = \left|F_{N-1}^{CPMG}(\omega,(N-1)x)\right|^2 + \left|F_1^{Hahn}(\omega,y)\right|^2 \\ + \underbrace{(-1)^{N-1} 2\Re\left\{e^{i\omega(TE-y)} F_{N-1}^{CPMG}(\omega,(N-1)x) \overline{F_1^{Hahn}(\omega,y)}\right\}}_{\text{Hahn/CPMG cross-term}}. \quad \text{(S.12)}$$

The analytical solution for the signal is then given by

$$M_{SDR}(TE,x,y,N) = M_{CPMG}((N-1)x, N-1) \times M_{\text{Hahn}}(y) \times M_{Cross}(TE,x,y,N). \quad \text{(S.13)}$$

The first two terms on the rigth-hand side of Eq. (S.13) are as in Eqs. (S.8) and (S.10), while the cross term $M_{Cross}(TE,x,y,N)$ has an argument

$$-\ln\{M_{Cross}\} = \Delta\omega_{SE}^2 \tau_c^2 \left[ \left(1 + e^{-\frac{y}{\tau_c}} - 2e^{-\frac{1}{2}\frac{y}{\tau_c}} - 2e^{\frac{1}{2}\frac{2x-y}{\tau_c}} + e^{\frac{x-y}{\tau_c}} + 4e^{\frac{1}{2}\frac{x-y}{\tau_c}} - 2e^{\frac{1}{2}\frac{x-2y}{\tau_c}} - 2e^{\frac{1}{2}\frac{x}{\tau_c}} + e^{\frac{x}{\tau_c}}\right) + \right.$$
$$(-1)^N \times \left( e^{-\frac{xN-x+y}{\tau_c}} - 2e^{-\frac{1}{2}\frac{2xN-2x+y}{\tau_c}} - 2e^{-\frac{1}{2}\frac{-4x+2xN+y}{\tau_c}} + e^{-\frac{-2x+xN+y}{\tau_c}} + 4e^{-\frac{1}{2}\frac{-3x+2xN+y}{\tau_c}} - 2e^{-\frac{1}{2}\frac{-3x+2xN+2y}{\tau_c}} + \right.$$
$$\left.\left. e^{-\frac{(N-1)x}{\tau_c}} - 2e^{-\frac{1}{2}\frac{x(-3+2N)}{\tau_c}} + e^{-\frac{x(N-2)}{\tau_c}} \right) \right] / \left( e^{\frac{x}{\tau_c}} + 1 \right). \quad \text{(S.14)}$$

With these general expressions at hand, one can consider the effects of a specific dynamics on DD; for example, the effects of free or a restricted diffusion. These will differ by their corresponding parameters $\Delta\omega_{SE}^2$ and $\tau_c$ and their respective spectral densities. In the restricted diffusion regime that concerns us we expect $x,y \gg \tau_c$; the main contribution to the overlap (S.3) causing the signal's decay, will then be dominated by the low frequencies at the center of the spectral density [12, 13]. In this case the signal decays of the sequences discussed in the previous parragraph become

$$M_{free}^{restricted}(TE) \approx \exp\left\{-\Delta\omega_{SE}^2 \tau_c (TE - \tau_c)\right\}, \quad \text{(S.15)}$$
$$M_{Hahn}^{restricted}(TE) \approx \exp\left\{-\Delta\omega_{SE}^2 \tau_c (TE - 3\tau_c)\right\}, \quad \text{(S.16)}$$
$$M_{CPMG}^{restricted}(TE,N) \approx \exp\left\{-\Delta\omega_{SE}^2 \tau_c (TE - (1+2N)\tau_c)\right\}, \quad \text{(S.17)}$$
$$M_{Cross}^{restricted}(TE,x,y,N) \approx \exp\left\{-\Delta\omega_{SE}^2 \tau_c^2\right\}. \quad \text{(S.18)}$$

and the overall SDR decay is given by

$$M_{SDR}^{restricted}(TE,x,y,N) \approx \exp\left\{-\Delta\omega_{SE}^2 \tau_c (TE - (2N+2)\tau_c) - \Delta\omega_{SE}^2 \tau_c^2\right\}, \quad \text{(S.19)}$$
$$= \exp\left\{-\Delta\omega_{SE}^2 \tau_c (TE - (1+2N)\tau_c)\right\} \quad \text{(S.20)}$$
$$= M_{CPMG}^{restricted}(TE,N). \quad \text{(S.21)}$$

Notice that in this restricted $x,y \gg \tau_c$ regime all SDR decays are equal, and are actually independent of the $x$ and $y$ values: for all cases particles have experienced a maximum displacement between pulses. By contrast, if $y \gg \tau_c$ but $x$ is much smaller than $\tau_c$, only the Hahn-echo portion of SDR will experience the restricted regime. In this limit $M_{SDR}^{y \gg \tau_c}(TE, x \sim 0, y, N) = M_{Hahn}^{restricted}(y) = \exp\left\{-\Delta\omega_{SE}^2 \tau_c (y - 3\tau_c)\right\}$. Important to highlight within the context of Fig. 2 that $M_{SDR}^{y \gg \tau_c}$ grows with an exponential-like behaviour as $x/\tau_c$ increasesuntil achieving the value $M_{SDR}^{y \gg \tau_c}(TE, x=y, y, N) = M_{CPMG}^{restricted}(TE,N) \approx \exp\left\{-\Delta\omega_{SE}^2 \tau_c (TE - (1+2N)\tau_c)\right\}$. Within this regime

$$M_{Cross-SDR}^{y \gg \tau_c}(TE,x,y,N) \approx \exp\left\{-\Delta\omega_{SE}^2 \tau_c^2 \left(1 - \frac{2e^{-\frac{x}{2\tau_c}}}{\left(1+e^{-\frac{x}{\tau_c}}\right)}\right)\left(1 + (-1)^N e^{-\frac{(N-1)x}{\tau_c}}\right)\right\}, \quad \text{(S.22)}$$

and the overall signal becomes

$$M_{SDR}^{y \gg \tau_c}(TE,x,y,N) \approx M_{Hahn}^{restricted}(y) \times M_{CPMG}[(N-1)x, (N-1)] \times M_{Cross-SDR}^{y \gg \tau_c}(x,N). \quad \text{(S.23)}$$

Equation (S.13) was used for fitting the experimental data of Fig. 5 of the main text; however, the approximate expression (S.23) is indistinguishable from Eq. (S.13) for our experimental parameters.

---

\end{widetext}\selectlanguage{american}%

\end{document}